\documentclass[12pt]{article}
\usepackage{graphicx}
\usepackage{amsmath}
\usepackage{amssymb}

\newcommand\pubnumber{}
\newcommand\pubdate{\today}

\def\epfl{Laboratoire de Physique des Hautes Energies\\
\'Ecole polytechnique f\'ed\'erale de Lausanne, CH-1015 Lausanne, Switzerland}

\def\Title#1{\begin{center} {\Large #1 } \end{center}}
\def\Author#1{\begin{center}{ \sc #1} \end{center}}
\def\Address#1{\begin{center}{ \it #1} \end{center}}

\newcommand\pubblock{\rightline{\begin{tabular}{l} \pubnumber\\
         \pubdate  \end{tabular}}}
\newenvironment{Abstract}{\begin{quotation}  }{\end{quotation}}
\newenvironment{Presented}{\begin{quotation} \begin{center} 
             PRESENTED AT\end{center}\bigskip 
      \begin{center}\begin{large}}{\end{large}\end{center} \end{quotation}}

\def\beq{\begin{equation}}
\def\eeq#1{\label{#1}\end{equation}}
\def\eeqn{\end{equation}}

\def\beqa{\begin{eqnarray}}
\def\eeqa#1{\label{#1}\end{eqnarray}}
\def\eeqan{\end{eqnarray}}

\let\bar=\overbar

\def\D{{\cal D}}

\def\Dslash{\not{\hbox{\kern-4pt $D$}}}
\def\dslash{\not{\hbox{\kern-2pt $\del$}}}

\def\msb{{\bar{\ssstyle M \kern -1pt S}}}

\RequirePackage{xspace}
\usepackage{relsize}
\def\to{\ensuremath{\rightarrow}\xspace}
\def\CP{\ensuremath{C\!P}\xspace}
\def\C{\ensuremath{C}\xspace}
\def\P{\ensuremath{P}\xspace}
\def\T{\ensuremath{T}\xspace}
\def\B{\ensuremath{B}\xspace}
\def\D{\ensuremath{D}\xspace}
\def\Db{\ensuremath{\bar{D}}\xspace}
\def\Dz{\ensuremath{D^0}\xspace}
\def\Dzb{\ensuremath{\bar{D}^0}\xspace}
\def\Dp{\ensuremath{D^+}\xspace}
\def\Ds{\ensuremath{D_s^+}\xspace}
\def\KS{\ensuremath{K^{0}_S}\xspace}

\def\mum{\ensuremath{\mu^-}\xspace}
\def\Kp{\ensuremath{K^+}\xspace}
\def\Km{\ensuremath{K^-}\xspace}
\def\pip{\ensuremath{\pi^+}\xspace}
\def\pim{\ensuremath{\pi^-}\xspace}

\def\ct{\ensuremath{C_T}\xspace}
\def\ctb{\ensuremath{\overline{C}_T}\xspace}
\def\at{\ensuremath{A_T}\xspace}
\def\atb{\ensuremath{\overline{A}_T}\xspace}
\def\atv{\ensuremath{a_{\CP}^{\T\text{-odd}}}\xspace}
\def\ap{\ensuremath{A_{\P}}\xspace}
\def\apb{\ensuremath{\overline{A}_{\P}}\xspace}
\def\ac{\ensuremath{A_{\C}}\xspace}
\def\acb{\ensuremath{\overline{A}_{\C}}\xspace}
\def\acp{\ensuremath{A_{\CP}}\xspace}
\def\acpb{\ensuremath{\overline{A}_{\CP}}\xspace}

\def\invfb{\ensuremath{\mathrm{fb}^{-1}}\xspace}

\newcommand{\tev}{\ensuremath{{\mathrm{\,Te\kern -0.1em V}}}\xspace}

\begin{document}
\begin{titlepage}
\pubblock

\vfill
\Title{\CP-violating triple-product asymmetries in Charm decays}
\vfill
\Author{ Maurizio Martinelli\\on behalf the LHCb and BaBar collaborations}
\Address{\epfl}
\vfill
\begin{Abstract}
The use of triple-product correlations is described in relation to the search for \CP violation in 4-body charm meson decays. 
The latest results from the LHCb and BaBar Collaborations are reported.
A novel interpretation of the asymmetries from triple-product correlations is used for the BaBar results, which enables the extraction of information on the properties of \D decays under parity and charge-conjugation transformations.
\end{Abstract}
\vfill
\begin{Presented}
8th International Workshop on the CKM Unitarity Triangle (CKM 2014)\\
Vienna, Austria, September 8--12, 2014
\end{Presented}
\vfill
\end{titlepage}
\def\thefootnote{\fnsymbol{footnote}}
\setcounter{footnote}{0}
%


\section{Introduction}

Violation of \CP symmetry in charm meson decays is expected to be extremely small in the Standard Model (SM)~\cite{Bianco:2003vb,Grossman:2006jg}, although recent calculations do not exclude effects up to a few times $10^{-3}$~\cite{Feldmann:2012js,Brod:2011re,Bhattacharya:2012ah}.
This small \CP violation effect in the SM leaves room for beyond-SM effects that, even if small, could produce an asymmetry significantly larger than that predicted from the SM.
The large samples of charm meson decays recorded at the \B factories, and more recently at LHCb, enable the search for \CP violation at the sub-percent level, hence approaching the largest SM predictions.

Three kinds of search are usually pursued in probing \CP violation. These involve  interference effects between the decay amplitudes, between the mixing and the decay amplitudes, and in the mixing amplitudes~\cite{Bianco:2003vb}.
In the present contribution, an alternative approach using triple-product correlations is shown.
If the invariant matrix element of a quasi-two-body decay is expressed in the terms of its scalar, vector and tensor contributing amplitudes, the asymmetry built from triple-product correlations is indeed found to be sensitive to \CP violation~\cite{Valencia:1988it}.
Furthermore, this observable has different sensitivity to strong phases than the direct \CP violation asymmetry observable.
In the first case, the observable is proportional to the cosine of the difference of the strong phases of the interfering amplitudes, while in the second case it is proportional to the sine of this angle.
Since in both cases these terms are multiplied by the sine of the difference between the \CP-violating weak phases, one is more sensitive to \CP violation when the difference of the strong phases of the interfering amplitudes is small, while the other is more sensitive when this difference is large~\cite{Datta:2003mj}.

\section{The LHCb measurement}

\begin{figure}[htb]
\centering
\includegraphics[width=0.35\textwidth]{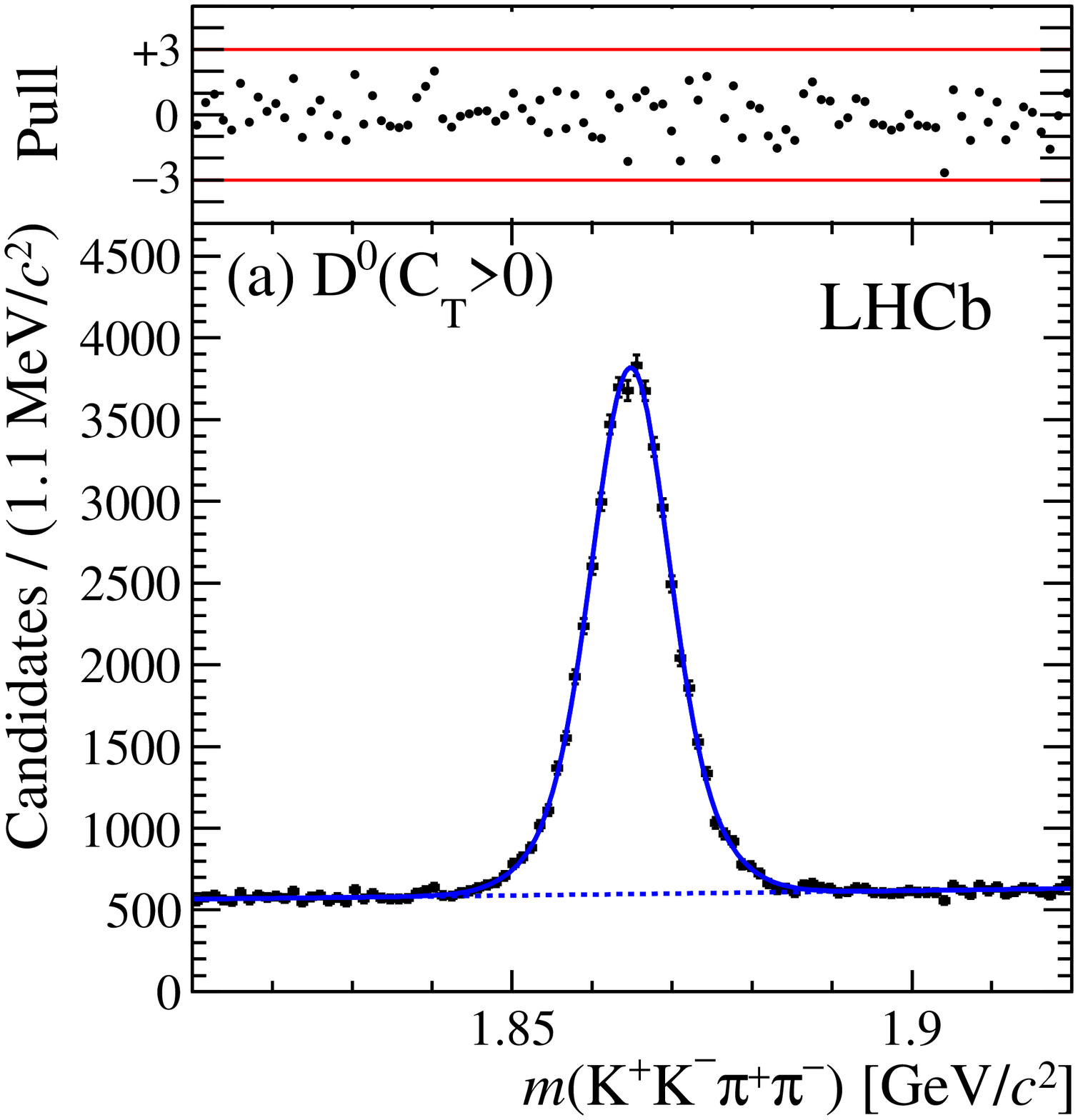}
\includegraphics[width=0.35\textwidth]{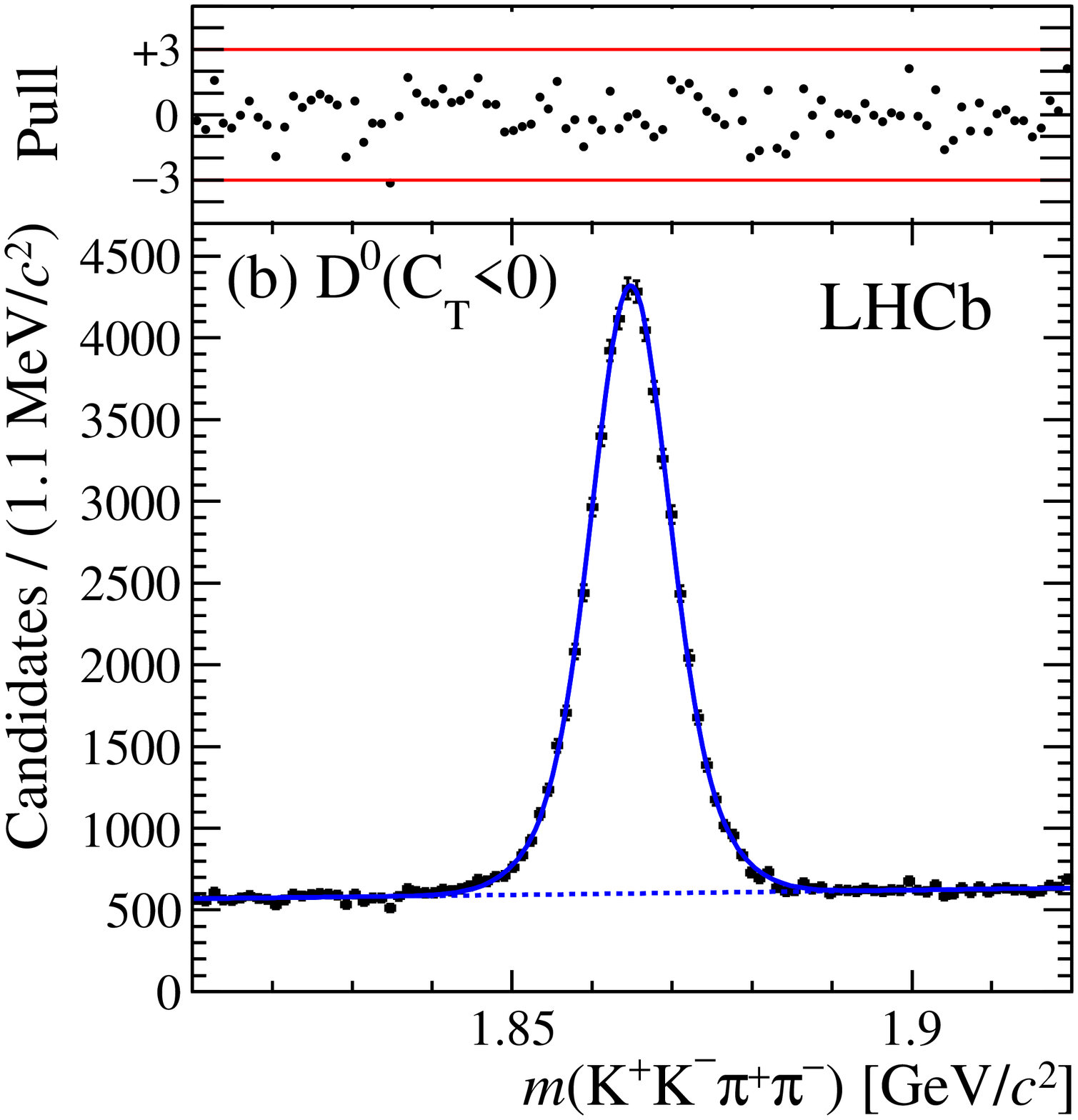}
\includegraphics[width=0.35\textwidth]{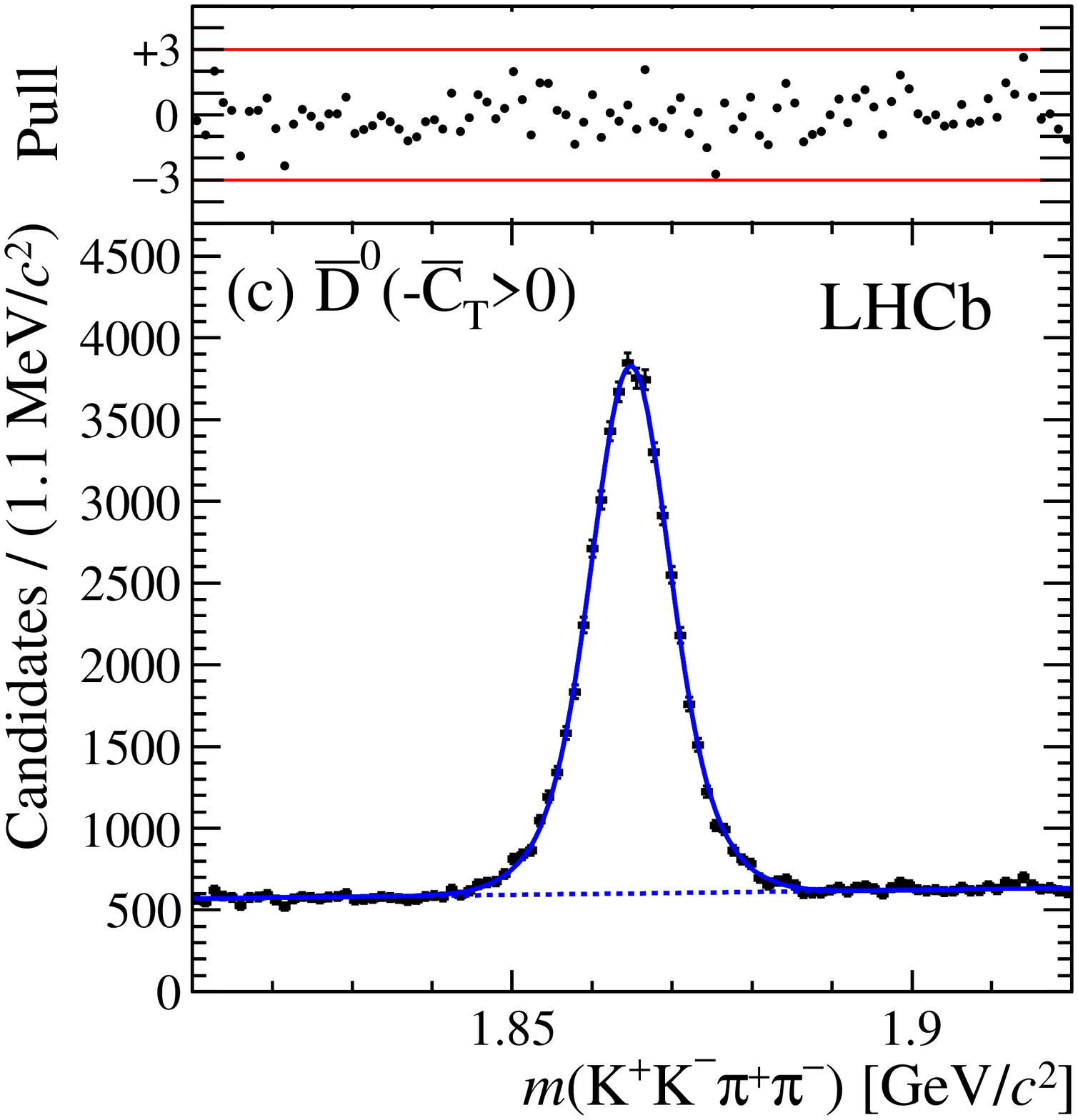}
\includegraphics[width=0.35\textwidth]{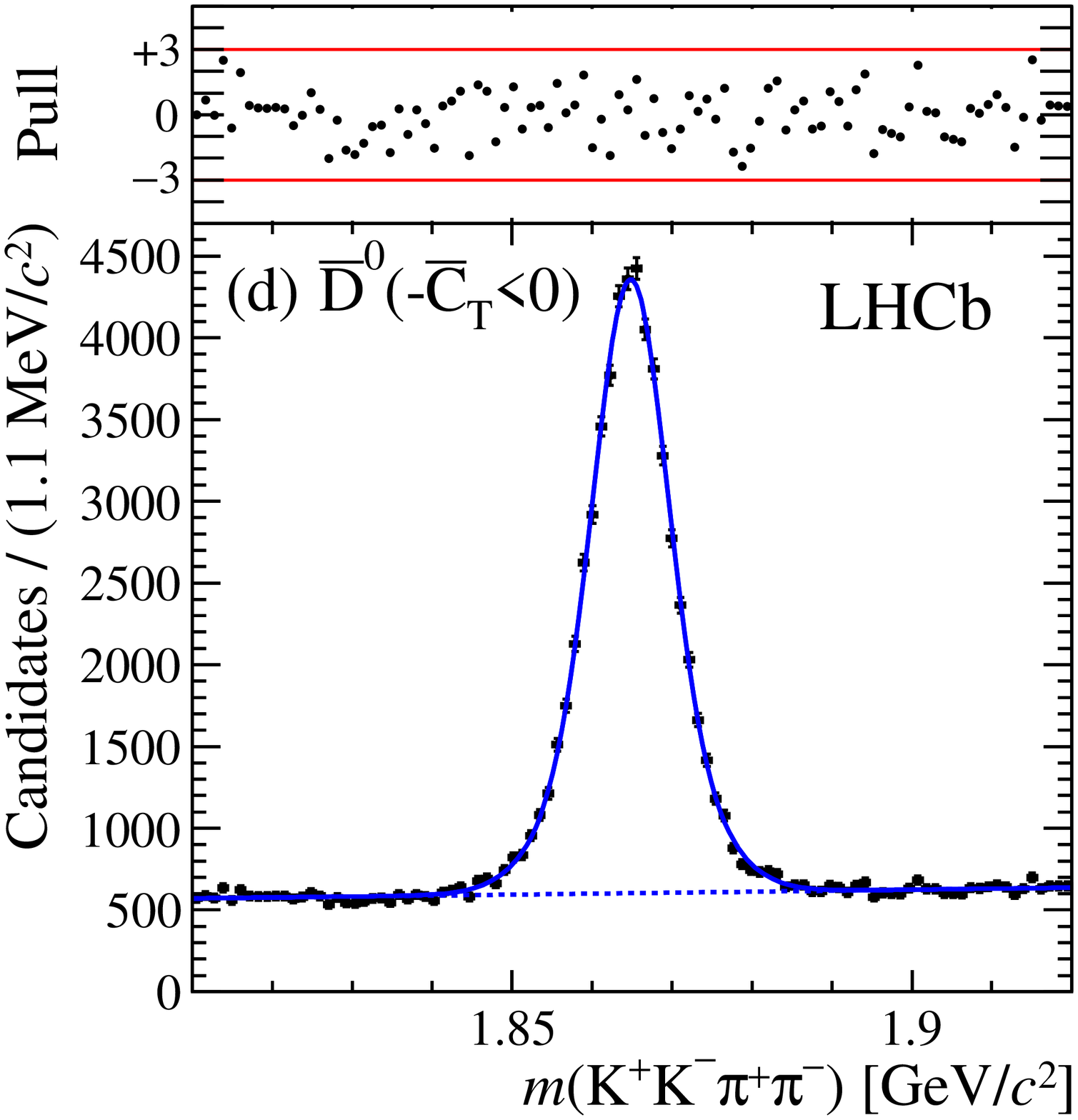}
\caption{\label{fig:d0massLHCb}\small\Dz-candidate mass distributions depending on flavour and the value of \ct~\cite{Aaij:2014qwa}.
In each case, the result of the fit is overlaid as a solid curve, while a dashed line represents the background. The distribution of normalised residual (pull), defined as the difference between the fit result and the data, divided by the data uncertainty, is shown above each data distribution.}
\end{figure}

The LHCb collaboration used triple-product correlations to search for \CP violation in $\Dz\to\Kp\Km\pip\pim$ decays\footnote{Throughout this document the use of charge conjugate reactions is implied, unless otherwise indicated.}~\cite{Aaij:2014qwa}.
The triple-product is defined by using the momenta of three out of the four daughters in the \Dz rest frame ($\ct\equiv\vec{p}_{\Kp}\cdot\vec{p}_{\pip}\times\vec{p}_{\pim}$).
A sample of 170,000 \Dz decays is found in 3 \invfb of data recorded by the LHCb detector in 2011 and 2012, when selecting them through partial reconstruction of semileptonic decays of the \B meson ($\B\to\Dz\mum X$, where $X$ indicates any system composed of charged and neutral particles).
The charge of the muon is used to identify the flavour of the \Dz meson.
The distributions of the \Dz meson candidates in four different regions defined by \Dz flavour and \ct value being greater or less then zero are shown in Figure~\ref{fig:d0massLHCb}.

The triple-product \ct is used to define the two asymmetries
\begin{align*}
\at = \frac{\Gamma(\Dz,\ct>0)-\Gamma(\Dz,\ct<0)}{\Gamma(\Dz,\ct>0)+\Gamma(\Dz,\ct<0)},\phantom{pp}
\atb = \frac{\Gamma(\Dzb,-\ctb>0)-\Gamma(\Dzb,-\ctb<0)}{\Gamma(\Dzb,-\ctb>0)+\Gamma(\Dzb,-\ctb<0)}.
\end{align*}
These are not \CP-violating asymmetries because they are proportional to the sine of the sum of the strong and weak phase differences~\cite{Valencia:1988it}.
The \CP violating observable is given by the difference of these two asymmetries $\atv\equiv (\at-\atb)/2$.

\begin{figure}[htb]
\centering
\includegraphics[width=0.32\textwidth]{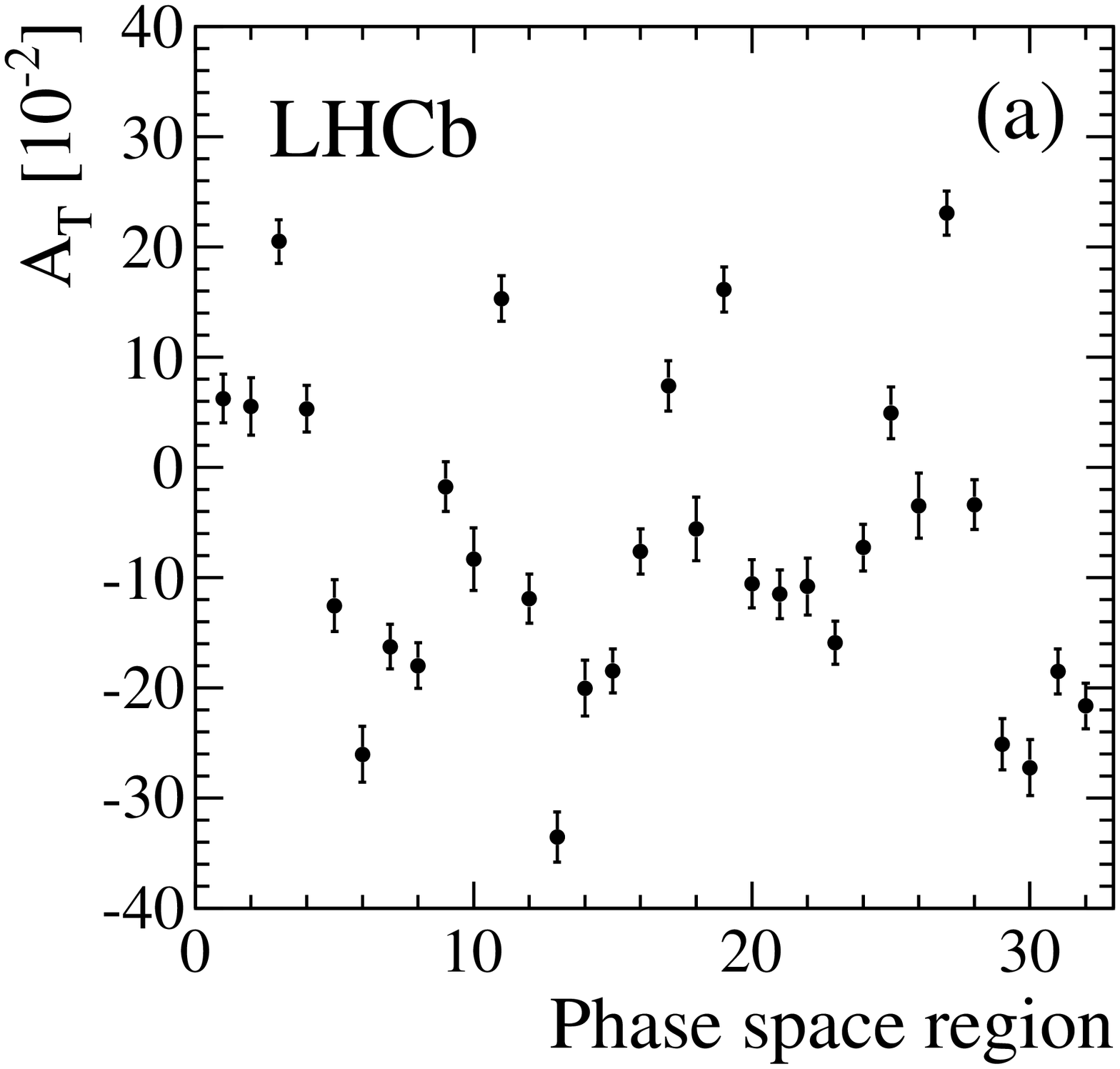}
\includegraphics[width=0.32\textwidth]{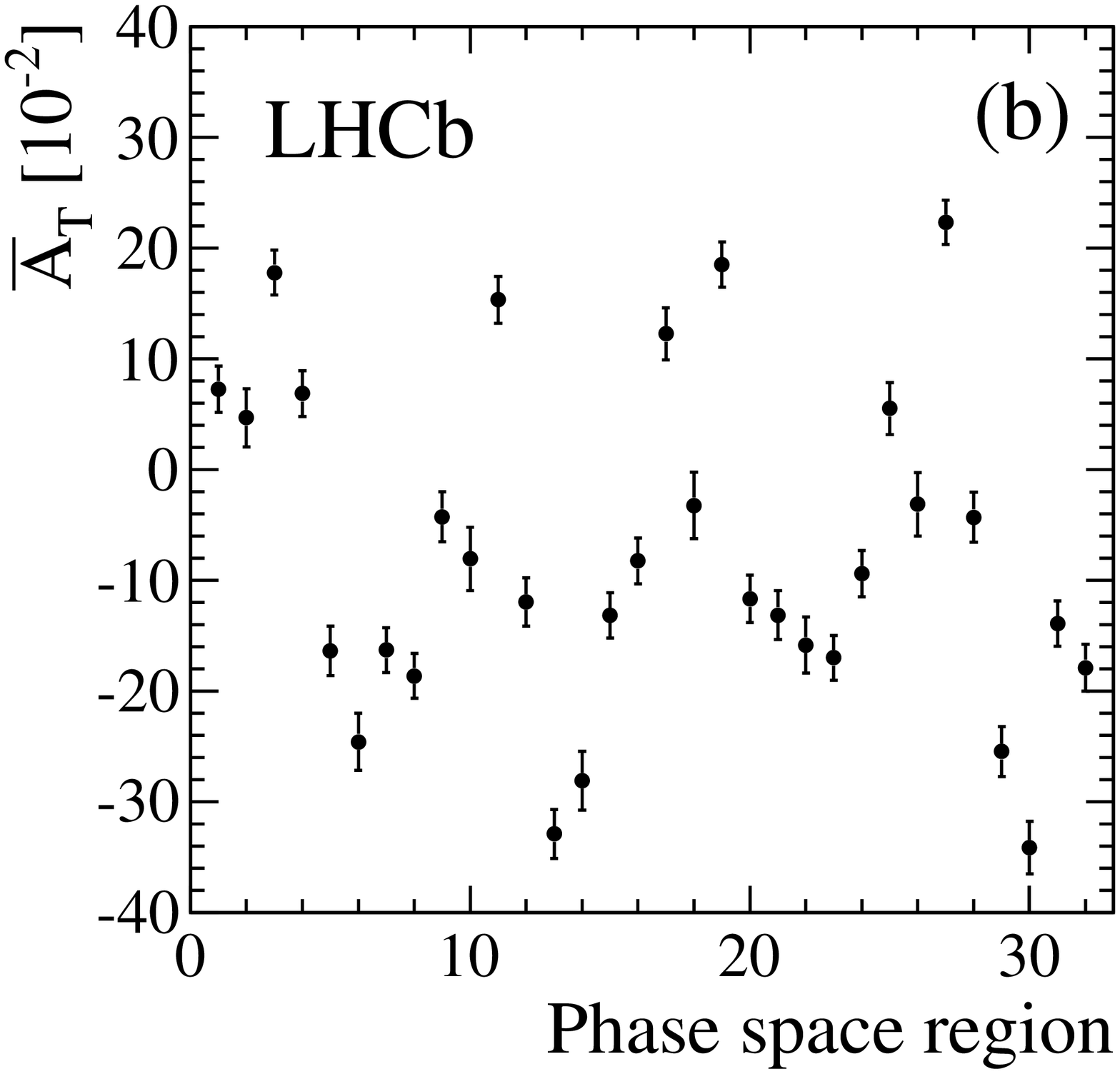}
\includegraphics[width=0.32\textwidth]{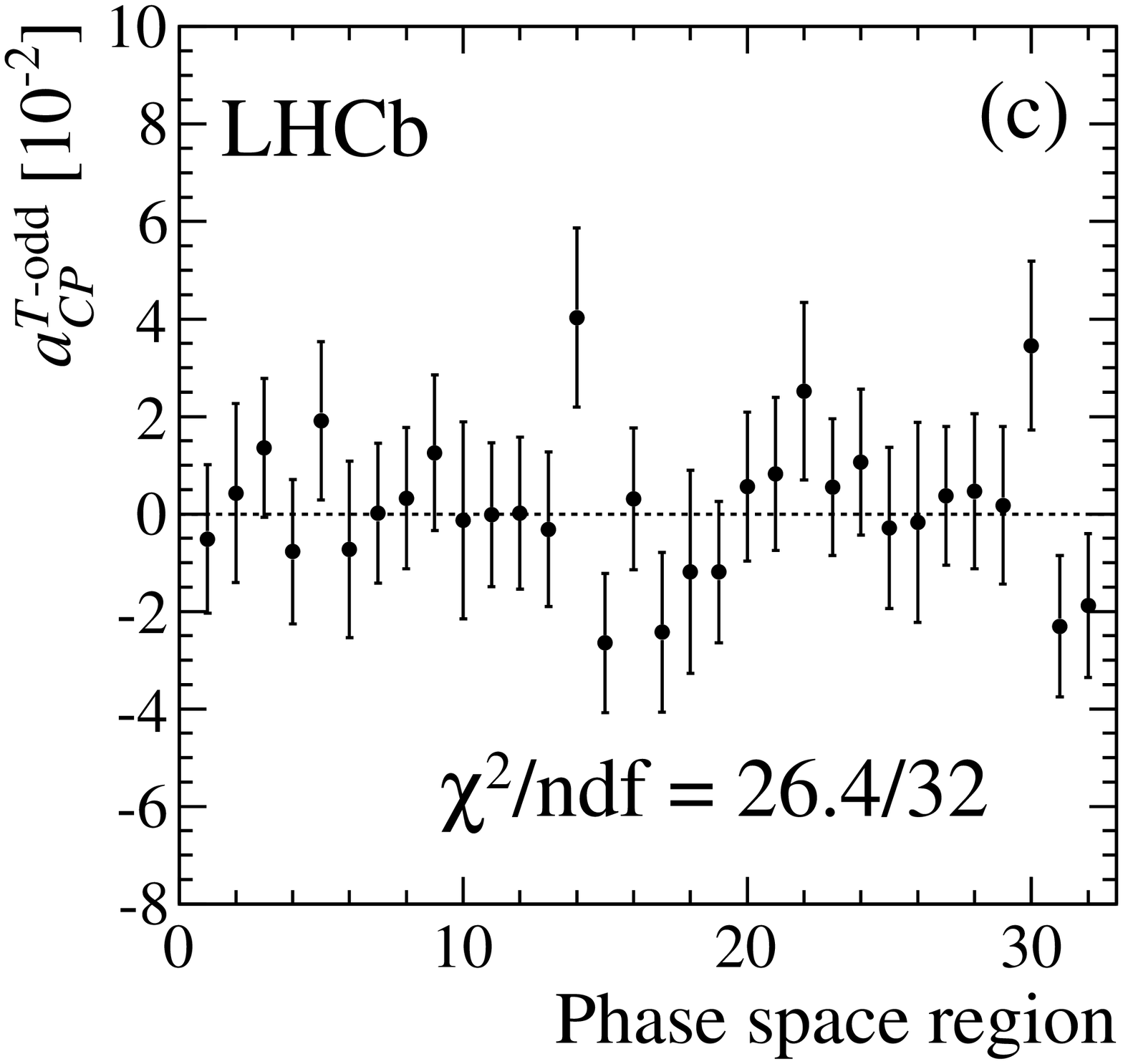}
\caption{\small\label{fig:asymPSLHCb}The \at (a), \atb (b), and \atv (c) asymmetries as a function of phase-space region~\cite{Aaij:2014qwa}.}
\end{figure}

\begin{figure}[htb]
\centering
\includegraphics[width=0.32\textwidth]{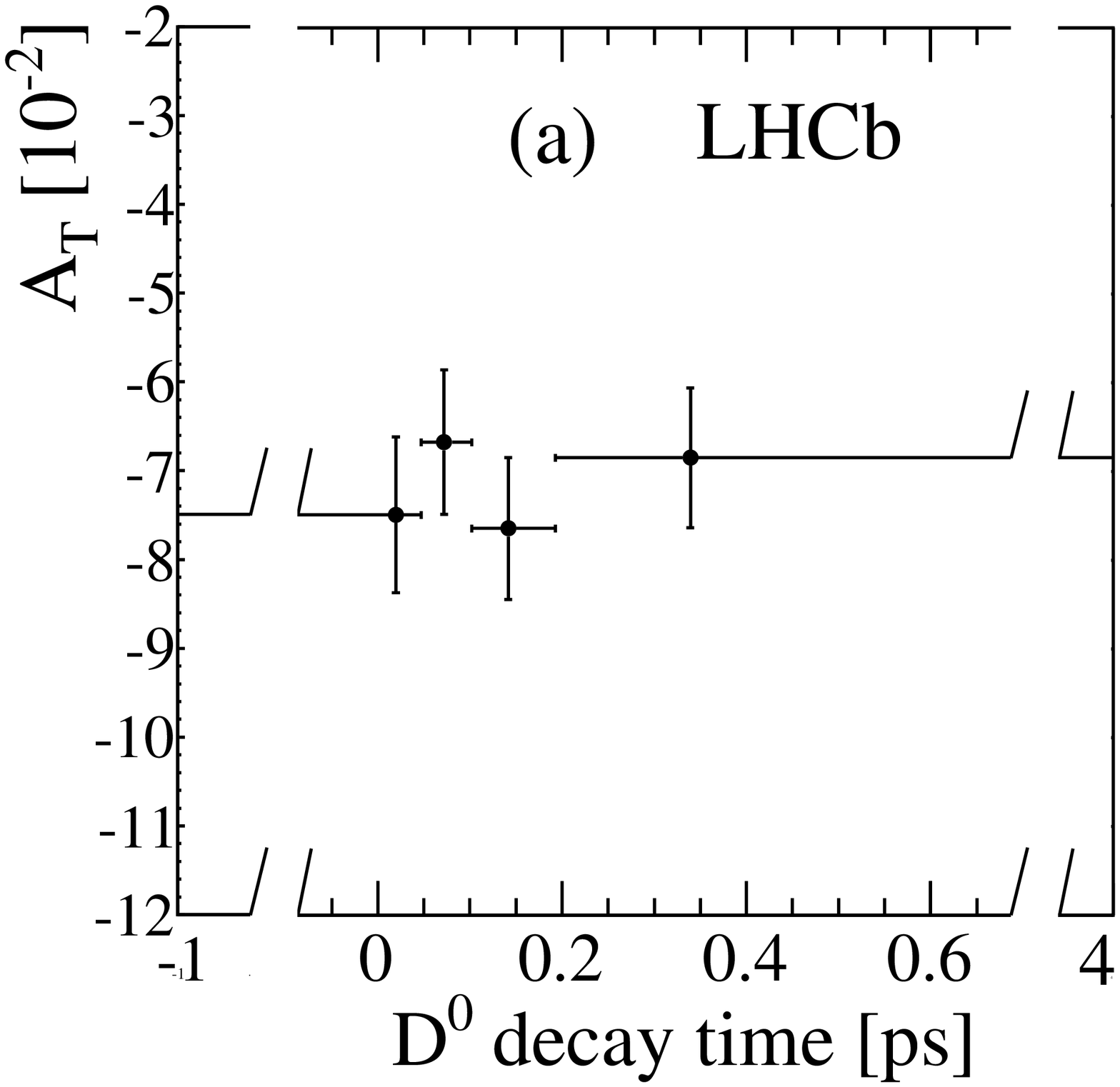}
\includegraphics[width=0.32\textwidth]{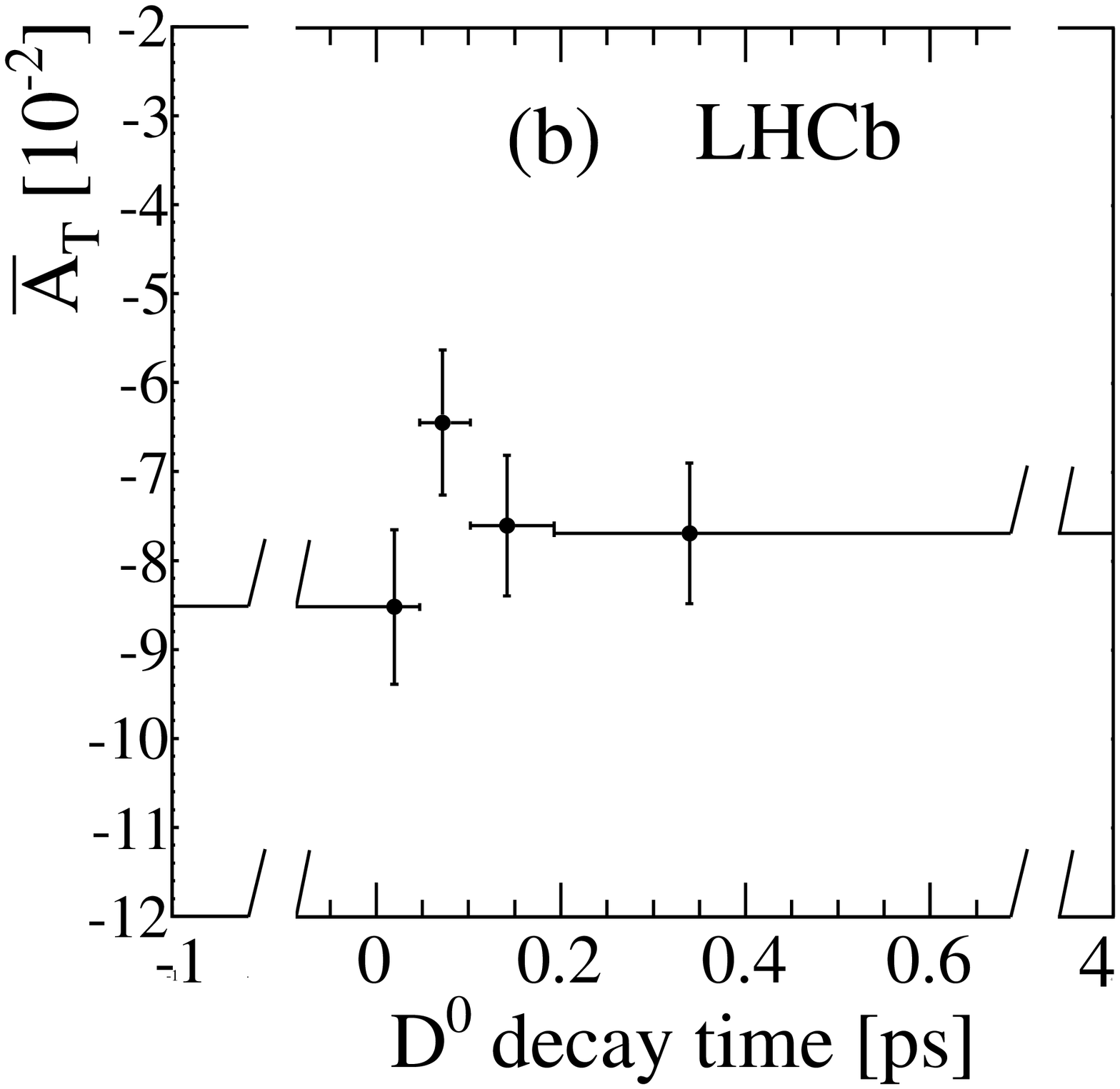}
\includegraphics[width=0.32\textwidth]{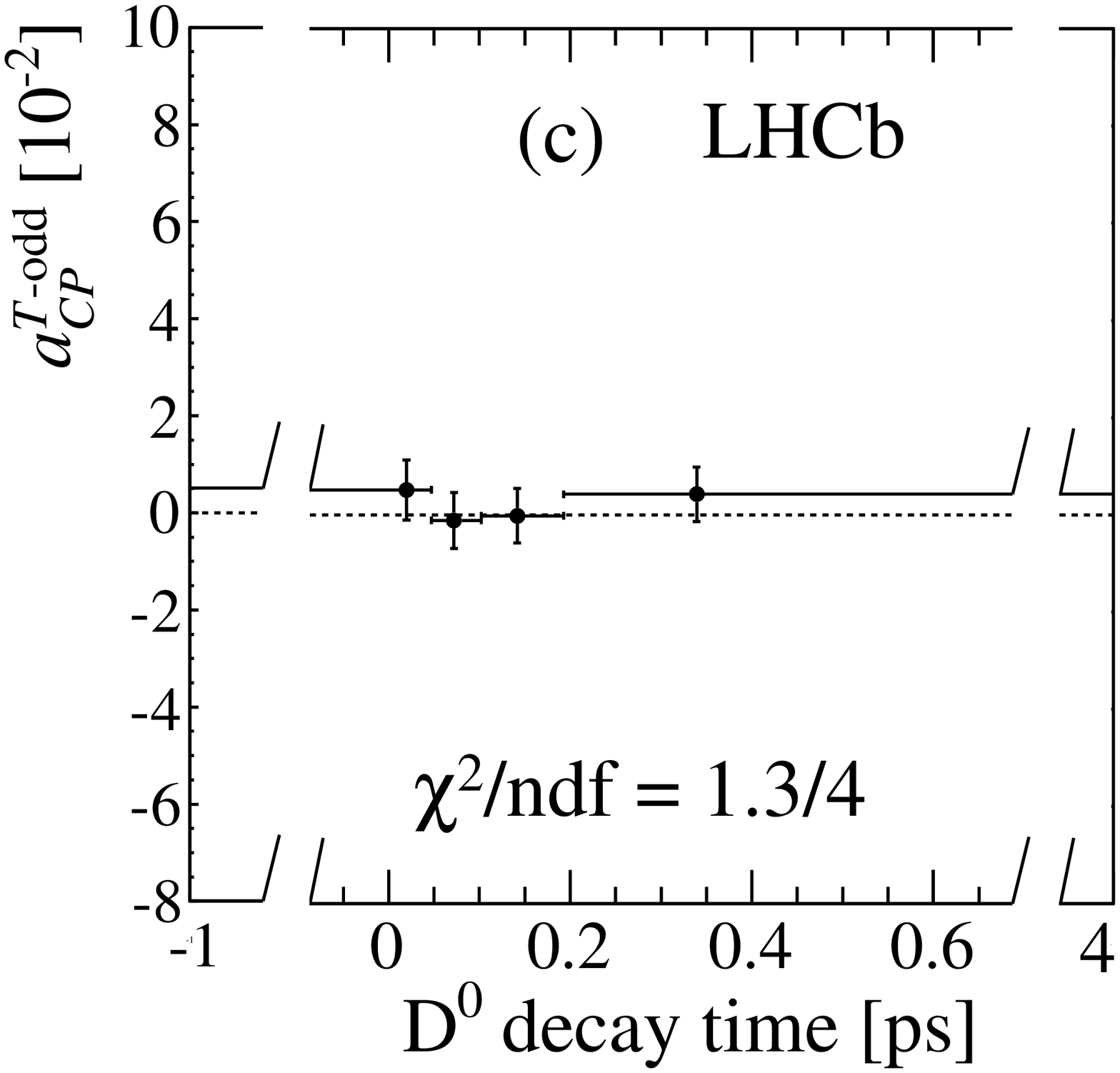}
\caption{\small\label{fig:asymDTLHCb}The \at (a), \atb (b), and \atv (c) asymmetries as a function of \Dz-decay-time region. The scale is broken for the first and last bin~\cite{Aaij:2014qwa}.}
\end{figure}

The size and the quality of the sample allowed the LHCb Collaboration to perform three different searches for \CP violation: a time-integrated measurement, a measurement as a function of phase-space region and a measurement as a function of \Dz-decay-time region, with the results shown in Figures~\ref{fig:asymPSLHCb} and \ref{fig:asymDTLHCb}.

The result of the integrated measurement is $\atv=(1.8\pm2.9(\text{stat})\pm0.4(\text{syst}))\times10^{-3}$.
No significant deviation from zero is observed either as a function of phase-space region or as a function of decay-time, and a $\chi^2$ probability test used to estimate the compatibility of the \CP conservation hypothesis gave the values 74\% and 83\%, for the respective distributions.

The non-\CP-violating asymmetries \at and \atb were also measured and found to be significantly different from zero, namely $\at=(-71.8\pm4.1(\text{stat})\pm1.3(\text{syst}))\times10^{-3}$ and $\atb=(-75.5\pm4.1(\text{stat})\pm1.2(\text{syst}))\times10^{-3}$. 
Furthermore, while their distributions are flat in \Dz decay time, they show significant deviations from the average values in different regions of the phase space, with local asymmetry values ranging from -30\% to 30\%.
These effects can be interpreted as being due to the various resonant contributions which produce different asymmetries as a result of final state interactions in different regions of the phase space.

An interesting feature of this measurement is that, by definition, the systematic uncertainties are very small.
The largest uncertainty on \atv is due to detector bias as estimated from a control sample of $\B\to\Dz(\Km\pip\pip\pim)\mum X$ decays. This is conservatively assigned as the statistical uncertainty of the measurement on this sample, since no significant bias is observed.
For the \at and \atb measurements, the background from prompt \Dz decays and flavour misidentification are the largest sources of uncertainty, but they cancel in \atv.

\section{A new interpretation of triple-product correlations}

A recent paper~\cite{Bevan:2014nva} suggests a novel interpretation of the triple-product correlation asymmetries previously mentioned, and suggests further asymmetries to investigate.
Since \ct is \P-odd, the \at and \atb asymmetries are interpreted as \P-violating asymmetries ($\ap=\at$ and $\apb=-\atb$).
Further, considering that $\C(\ap)=\apb$ and $\CP(\ap)=-\apb$, one can build the \C- and \CP-violating asymmetries $a_{\C}^{\P}\equiv(\ap-\apb)/2$ and $a_{\CP}^{\P}\equiv(\ap+\apb)/2=\atv$.

The same exercise can be repeated for \C and \CP.
Considering that $\C(\ct)=\ctb$, one can define the \C-violating asymmetries
\begin{align*}
\ac = \frac{\Gamma(\Dzb,\ctb<0)-\Gamma(\Dz,\ct<0)}{\Gamma(\Dzb,\ctb<0)+\Gamma(\Dz,\ct<0)},\phantom{pp}
\acb = \frac{\Gamma(\Dzb,\ctb>0)-\Gamma(\Dz,\ct>0)}{\Gamma(\Dzb,\ctb>0)+\Gamma(\Dz,\ct>0)}.
\end{align*}
Since $\P(\ac)=\acb$ and $\CP(\ac)=-\acb$, $a_{\P}^{\C}\equiv(\ac-\acb)/2$ and $a_{\CP}^{\C}\equiv(\ac+\acb)/2$ are \P- and \CP-violating asymmetries, respectively.

Finally, since $\CP(\ct)=-\ctb$, the \CP-violating asymmetries
\begin{align*}
\acp = \frac{\Gamma(\Dzb,\ctb>0)-\Gamma(\Dz,\ct<0)}{\Gamma(\Dzb,\ctb>0)+\Gamma(\Dz,\ct<0)},\phantom{pp}
\acpb = \frac{\Gamma(\Dzb,\ctb<0)-\Gamma(\Dz,\ct>0)}{\Gamma(\Dzb,\ctb<0)+\Gamma(\Dz,\ct>0)}.
\end{align*}
can be defined.
Applying \P and \C to these asymmetries $\P(\acp)=\acpb$ and $\C(\acp)=-\acpb$, and the \P- and \C-violating asymmetries $a_{\P}^{\CP}\equiv(\acp-\acpb)/2$ and $a_{\C}^{\CP}\equiv(\acp+\acpb)/2$ are obtained.

\section{Reinterpretation of BaBar results}

The BaBar Collaboration reinterpreted their previous results on \CP violation using triple-product correlations in prompt $\Dz\to\Kp\Km\pip\pim$~\cite{delAmoSanchez:2010xj} and $D_{(s)}^+\to\Kp\KS\pip\pim$~\cite{Lees:2011dx} decays in light of Ref.~\cite{Bevan:2014nva}.
The \at and \atb asymmetries previously measured are translated into yields following $\Gamma(\D,\ct\gtrless0) = N_{\D}(1\pm\ap)$ and $\Gamma(\bar\D,\ctb\gtrless 0) = N_{\Db}(1\mp\apb)$, and the event rates in Table~\ref{tab:yieldsBaBar} are obtained.

\begin{table}[ht]
\begin{center}
\begin{tabular}{c|ccc}  
Event &  \Dz &  \Dp & \Ds\\ \hline
$\Gamma(\D,\ct>0)$ & $10974\pm117$  & $5406\pm136$ &  $6792\pm135$  \\
$\Gamma(\D,\ct<0)$ & $12587\pm125$  & $5287\pm131$ &  $8287\pm131$  \\
$\Gamma(\bar\D,\ctb>0)$ & $12380\pm124$  & $5073\pm104$ &  $7886\pm121$  \\
$\Gamma(\bar\D,\ctb<0)$ & $10749\pm116$  & $5443\pm111$ &  $6826\pm107$  \\ \hline
Total & $46690\pm241$  & $21209\pm242$ &  $29791\pm248$  \\
\end{tabular}
\caption{\small\label{tab:yieldsBaBar}Number of signal events extracted from the asymmetry measurements in previous BaBar publications~\cite{delAmoSanchez:2010xj,Lees:2011dx}.}
\end{center}
\end{table}

\begin{table}[htb]
\begin{center}
\begin{tabular}{c|ccc}  
Asymmetry &  $\Dz\to\Kp\Km\pip\pim$ &  $\Dp\to\Kp\KS\pip\pim$ & $\Ds\to\Kp\KS\pip\pim$\\ \hline
\ap  & $-6.9\pm0.7\pm0.6 (7.5)\phantom{1}$  & $\phantom{-}1.1\pm1.4\pm0.6 (0.7)$ &  $-9.9\pm1.1\pm0.8 (7.3)$\\
\apb & $\phantom{-}7.1\pm0.7\pm0.4 (8.8)\phantom{1}$  & $-3.5\pm1.4\pm0.7 (2.2)$ &  $\phantom{-}7.2\pm1.1\pm1.1 (4.6)$\\
$a_{\C}^{\P}$ & $-7.0\pm0.5\pm0.1 (13.5)$  & $\phantom{-}2.3\pm1.1\pm0.2 (2.1)$ &  $-8.6\pm0.9\pm0.2 (9.3)$\\
$a_{\CP}^{\P}$& $\phantom{-}0.1\pm0.5\pm0.4 (0.2)\phantom{1}$  & $-1.2\pm1.0\pm0.5 (1.1)$ &  $-1.4\pm0.8\pm0.3 (1.6)$\\
\ac  & $\phantom{-}6.0\pm0.7\pm0.1 (8.3)\phantom{1}$  & $-2.6\pm1.6\pm0.5 (1.6)$ &  $\phantom{-}8.0\pm1.3\pm0.5 (5.7)$\\
\acb & $-7.9\pm0.7\pm0.1 (10.8)$  & $\phantom{-}2.0\pm1.6\pm0.5 (1.2)$ &  $-9.2\pm1.2\pm0.5 (7.1)$\\
$a_{\P}^{\C}$ & $\phantom{-}7.0\pm0.5\pm0.1 (13.5)$  & $-2.3\pm1.1\pm0.2 (2.1)$ &  $\phantom{-}8.6\pm0.9\pm0.2 (9.3)$\\
$a_{\CP}^{\C}$& $-0.9\pm0.5\pm0.0 (1.8)\phantom{1}$  & $-0.4\pm1.1\pm1.0 (0.3)$ &  $-0.6\pm0.9\pm1.0 (0.4)$\\
\acp  & $-0.8\pm0.7\pm0.4 (1.0)\phantom{1}$  & $-1.6\pm1.6\pm0.8 (0.9)$ &  $-2.0\pm1.2\pm0.8 (1.4)$\\
\acpb & $-1.0\pm0.8\pm0.4 (1.1)\phantom{1}$  & $\phantom{-}0.8\pm1.6\pm0.8 (0.5)$ &  $\phantom{-}0.8\pm1.3\pm0.9 (0.5)$\\
$a_{\P}^{\CP}$ & $\phantom{-}0.1\pm0.5\pm0.4 (0.2)\phantom{1}$  & $-1.2\pm1.1\pm0.6 (1.0)$ &  $-1.4\pm0.9\pm0.6 (1.3)$\\
$a_{\C}^{\CP}$& $-0.9\pm0.5\pm0.0 (1.8)\phantom{1}$  & $-0.4\pm1.1\pm1.0 (0.3)$ &  $-0.6\pm0.9\pm1.0 (0.4)$\\
\end{tabular}
\caption{\small\label{tab:asymsBaBar}Asymmetry measurements extracted from the BaBar data (percent values). The first uncertainty is statistical, the second systematic. The statistical significance of the asymmetry is quoted in parentheses.}
\end{center}
\end{table}

The systematic uncertainties are extracted following the same considerations and assuming them to be distributed as Gaussian functions.
A few specific sources of systematic uncertainty are considered in addition to those identified in the original publications~\cite{delAmoSanchez:2010xj,Lees:2011dx}. 
The slow pion reconstruction asymmetry, which affects \ac and \acp asymmetries from \Dz decays, is found to be negligible.
Neutral kaon regeneration and interference introduces an effect of 0.06\%~\cite{delAmoSanchez:2011zza}.
The asymmetry between of \Kp- and \Km-interactions in the BaBar detector material introduces an uncertainty of 0.5\%.
Both uncertainties affect only the \Dp and \Ds decay mode.

Table~\ref{tab:asymsBaBar} shows the results for all the aforementioned asymmetries and all the decay modes considered.
No evidence of \CP violation is found.
The tests performed on the \Dp show no significant asymmetry.
A significant deviation from 0, indicating violation of parity and charge-conjugation (\ap, \apb, $a_{\P}^{\C}$, $a_{\C}^{\P}$, \ac, \acb), is found for the \Dz and \Ds decays.
Similar effects are not observed in \Dp decays. 
A better understanding of the amplitude structure of these decays and theoretical input are needed to fully understand this difference.

\section{Conclusions}

The triple-product correlations provide alternative and complementary measurements with which to search for \CP violation in multi-body particle decays.
Recent studies suggest that these correlations can be used to probe \C and \P symmetries as well.
Both LHCb and BaBar Collaborations searched for \CP violation using \T-odd correlations in $\Dz\to\Kp\Km\pip\pim$ decays, finding $\atv\text{(LHCb)}=(1.8\pm2.9(\text{stat})\pm0.4(\text{syst}))\times10^{-3}$ and $\atv\text{(BaBar)}=(1.0\pm5.1(\text{stat})\pm4.4(\text{syst}))\times10^{-3}$.
The two results are consistent with each other and show no sign of \CP violation.
Nevertheless, they demonstrated that this observable is affected by systematic uncertainties which are  very small, and therefore is appropriate for the study of very large data samples expected at LHCb after LHC Run 2 ($\sim10$ \invfb) or at future experiments, such as Belle-II and the LHCb Upgrade.

\end{document}